\documentstyle[pra,aps,tighten,eqsecnum,multicol]{revtex}

\title{Universal quantum computation using only projective
measurement, quantum memory, and preparation of the $|0\rangle$ state}

\author{Michael A. Nielsen\thanks{nielsen@physics.uq.edu.au}
\vspace{3mm} \\
Centre for Quantum Computer Technology, 
University of Queensland, 
Queensland 4072, Australia.} 

\date{\today}

\begin{document}

\pagestyle{plain}
\pagenumbering{arabic}

\maketitle


\begin{multicols}{2}[]
\narrowtext

%
%

{\bfseries What resources are universal for quantum computation?  In
the standard model, a quantum computer consists of a sequence of
unitary gates acting coherently on the qubits making up the computer.
This paper shows that a very different model involving only projective
measurements, quantum memory, and the ability to prepare the
$|0\rangle$ state is also universal for quantum computation.  In
particular, {\itshape no coherent unitary dynamics are involved in the
computation}.}

%
%

Recall that in the standard quantum circuits model (see, for
example,~\cite{Nielsen00a,Preskill98c} for a review) a quantum
computation consists of three elements: (a) an {\em initialization
stage} in which the computer is prepared in an $n$-qubit computational
basis state; (b) a {\em processing stage} in which a sequence of one-
and two-qubit unitary gates is applied to the computer; and (c) a {\em
read-out stage} in which the result of the computation is read out by
measuring some subset of the qubits in the computational basis.

%

This paper demonstrates that a surprising alternate model is also
universal for quantum computation.  In this {\em measurement model}
for quantum computation only the following three operations are
allowed: (a) preparation of qubits in the $|0\rangle$ state; (b)
storage of qubits (quantum memory); and (c) (non-destructive)
projective measurements on up to four qubits at a time, in an
arbitrary basis.  What is surprising about the universality of this
model is that no coherent dynamical operations are allowed, contrary
to the widespread folklore belief that such operations are crucial to
universal quantum computation.

%
%

Substantial prior work has been done on the physical requirements for
universal quantum computation.  Barenco {\em et al}~\cite{Barenco95a}
showed that controlled-{\sc not} and single-qubit unitary operations
are sufficient to do universal quantum computation.  Deutsch, Ekert
and Barenco~\cite{Deutsch95a} and Lloyd~\cite{Lloyd95a} showed that
{\em almost any} two-qubit quantum gate is universal.  More recently,
Dodd {\em et al}~\cite{Dodd01a} have shown that {\em any} two-body
Hamiltonian entangling qubits, together with local unitary operations,
is universal for computation.  (See also related work
in~\cite{Wocjan01a,Janzing01a,Bennett01a,Leung01a}.)

%
%

Thus, the conventional approach to universality has been to identify a
set of coherent dynamical operations universal for quantum
computation.  Recently, however, it has been realized that {\em
quantum measurement} is a powerful primitive element that can be
performed during a computation.  For example, Raussendorf and
Briegel~\cite{Raussendorf01a} have shown that by combining quantum
measurements with the ability to prepare a special ``cluster'' state
using a coherent Ising-type interaction, it is possible to do
universal quantum computation.  Knill, Laflamme and
Milburn~\cite{Knill01a} have shown that single-photon detection and
single-photon sources enable universal quantum computation in optics,
using otherwise relatively easy coherent dynamical operations from
linear optics.

%
%

In this paper I extend these results to show that no coherent dynamics
at all are required to do universal quantum computation.  The results
of this paper build on a method developed by Nielsen and
Chuang~\cite{Nielsen97c}, who showed how to stochastically
``teleport'' a quantum gate from one location to another.  Elegant
generalizations of this method have been developed
in~\cite{Vidal00d,Vidal01a,Huelga00a,Huelga01a}.  The method has also
been applied by Gottesman and Chuang~\cite{Gottesman99a} to develop
fault-tolerant constructions for quantum gates, and further
work~\cite{Zhou00a} has been done simplifying such constructions.
Note also that the optical quantum computer proposed by Knill,
Laflamme and Milburn~\cite{Knill01a} is based on the gate
teleportation idea.

%
%

In order to show that the measurement model can simulate the standard
model of quantum computation we need only show that the measurement
model can simulate the controlled-{\sc not} gate, as well as any
single-qubit unitary gate.  We begin by explaining how the measurement
model can simulate the action of an arbitrary single-qubit gate $U$ on
a single-qubit state $|\psi\rangle$.

%
%

The first step is to use the measurement model to prepare one of the
two-qubit states $|U_j\rangle$ defined by $|U_j\rangle \equiv (I
\otimes U\sigma_j) (|00\rangle+|11\rangle)/\sqrt 2$, where $\sigma_j$
are the four Pauli $\sigma$ matrices, $I, \sigma_x, \sigma_y$, and
$\sigma_z$.  Simple algebra shows that these states form an
orthonormal basis for the state space of two qubits, so we can achieve
this state preparation by first preparing the state $|00\rangle$, and
then measuring in the orthonormal basis of states $|U_j\rangle$.  It
is important to note that the construction presented below works no
matter which of the four states $|U_j\rangle$ is output from this
measurement procedure.

%
%

Having prepared the state $|U_j\rangle$ offline, we now attempt to use
this state and Bell-basis measurements to perform the operation $U$ on
$|\psi\rangle$.  The basic operation required to achieve this is the
{\em gate teleportation} circuit shown in
Fig.~\ref{fig:basic_circuit}.  Following~\cite{Nielsen97c}, to analyse
the output from the third line of this circuit, it is convenient to
note that $U\sigma_j$ acting on the third qubit commutes with the Bell
measurement on the first two qubits, so Fig.~\ref{fig:basic_circuit}
is equivalent to Fig.~\ref{fig:easy_circuit}.  The circuit in
Fig.~\ref{fig:easy_circuit} is easy to understand; except for the
final gate $U \sigma_j$ acting on the third line, it is just quantum
teleportation~\cite{Bennett93a}, and thus the final state output from
the third line is $U \sigma_j \sigma_m |\psi\rangle$, where $\sigma_m$
is one of the four Pauli matrices determined by the measurement result
$m$ from the Bell measurement.  Note that each of the four possible
outcomes $U\sigma_j|\psi\rangle, U\sigma_j \sigma_x |\psi\rangle,
U\sigma_j \sigma_y |\psi\rangle$ and $U\sigma_j \sigma_z |\psi\rangle$
occurs with equal probability $1/4$.

\begin{figure}[t]
\begin{center}
\setlength{\unitlength}{1cm}
\begin{picture}(6.1,4)
\put(0.15,2.9){$|\psi\rangle$}		
\put(-0.2,1.45){$|U_j\rangle$}		
\put(0.5,1.42){$\Bigg\{$}		
\put(0.8,3){\line(1,0){1}}		
\put(0.8,2){\line(1,0){1}}		
\put(0.8,1){\line(1,0){5}}		
\put(1.8,1.5){\framebox(2.5,2){}}	
\put(2.75,2.9){Bell}			
\put(2,1.9){Measurement}		
\put(4.3,2.55){\line(1,0){0.5}}		
\put(4.3,2.45){\line(1,0){0.5}}		
\put(4.9,2.45){$m$}  			
\put(5.9,0.9){?}			
\end{picture}
\end{center}
\caption{Circuit to teleport the gate $U$ 
in the measurement model of quantum computation.  The state
$|U_j\rangle$ is prepared offline. $m = 0,1,2,3$ is the outcome of the
measurement in the Bell basis. \label{fig:basic_circuit}} 
\end{figure}
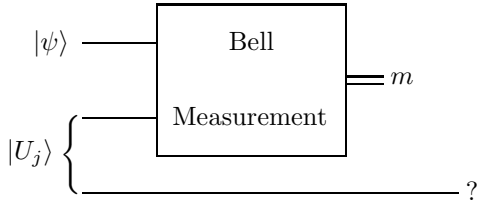

\begin{figure}[t]
\begin{center}
\setlength{\unitlength}{1cm}
\begin{picture}(7,4)
\put(0.8,2.9){$|\psi\rangle$}		
\put(-0.1,1.45){$\frac{|00\rangle+
	|11\rangle}{\sqrt 2}$}		
\put(1.2,1.42){$\Bigg\{$}		
\put(1.5,3){\line(1,0){1}}		
\put(1.5,2){\line(1,0){1}}		
\put(1.5,1){\line(1,0){3.9}}		
\put(2.5,1.5){\framebox(2.5,2){}}	
\put(3.45,2.9){Bell}			
\put(2.7,1.9){Measurement}		
\put(5,2.55){\line(1,0){0.5}}		
\put(5,2.45){\line(1,0){0.5}}		
\put(5.6,2.45){$m$}  			
\put(5.4,0.5){\framebox(1,1){$U\sigma_j$}}	
\put(6.4,1){\line(1,0){0.5}}		
\put(7,0.9){?}				
\end{picture}
\end{center}

\caption{This circuit produces the same output as
Fig.~\ref{fig:basic_circuit}, and is introduced merely to ease the
analysis of Fig.~\ref{fig:basic_circuit}.  Note that not all the
elements depicted in this circuit are allowed in the measurement model
of quantum computation, while those in Fig.~\ref{fig:basic_circuit}
are.\label{fig:easy_circuit}} \end{figure}
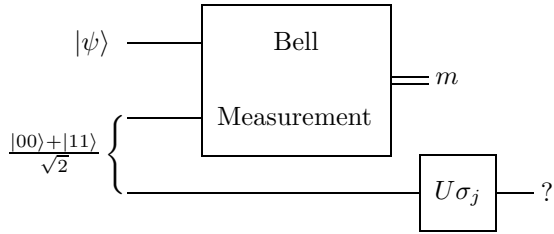

%
%

Thus, with probability $1/4$ the result $m$ is the same as $j$, and
the gate teleportation succeeds, with the gate $U$ being applied to
the qubit.  With probability $3/4$, however, $m \neq j$, and the
incorrect operation $U \sigma_j \sigma_m$ is applied to the qubit.  At
first glance this appears to create a significant problem, since we
can't just discard the qubit and start again, as it may have been in
an unknown state produced as part of a quantum computation.
Fortunately, this problem can be avoided by teleporting the gate $U
\sigma_m \sigma_j U^{\dagger}$.  If this succeeds, which occurs with
probability $1/4$, then the net action is $U \sigma_m \sigma_j
U^{\dagger} U \sigma_j \sigma_m = U$, as desired.  If the second gate
teleportation fails, then the net action is $U \sigma_m \sigma_j
U^{\dagger} \sigma_{m'} \sigma_{j'} U \sigma_j \sigma_m$, where
$|U_{j'}\rangle$ is the state that was prepared to do the second gate
teleportation, and $m'$ is the measurement result from the second gate
teleportation.  This is still okay, as we can use a similar gate
teleportation procedure again to attempt to obtain the correct
dynamics $U$.  In general, the procedure we use to simulate $U$ in the
measurement model is as follows:

\newcounter{program}
\begin{list}{\bf \arabic{program}} {\usecounter{program}
\setlength{\parsep}{0cm} \setlength{\itemsep}{0cm}
\setlength{\labelsep}{0.5cm} } 
\item {\bf Initialization} 
\item \hspace{0.5cm} $r := 1$ 
	\hfill{; loop counter}
\item \hspace{0.5cm} $C_0 := I$ 
	\hfill{; cumulative effect after $0$ iterations}
\item {\bf Main Loop} 
\item \hspace{0.5cm} $A^r := U C_{r-1}^{\dagger}$ 
	\hfill{; the gate we attempt to teleport}
\item \hspace{0.5cm} Prepare $|A^r_{j_r}\rangle$
\item \hspace{0.5cm} Teleport $A^r$, return result $m_r$ 
\item \hspace{0.5cm} $C_r := A^r \sigma_{j_r} \sigma_{m_r} C_{r-1}$
	\hfill{; cumulative effect}
\item \hspace{0.5cm} {\bf Case:} $m_r = j_r$
	\hfill{; success}
\item \hspace{1cm} {\bf Halt} 
	\hfill{; stop}
\item \hspace{0.5cm} {\bf Case:} $m_r \neq j_r$ 
	\hfill{; failure}
\item \hspace{1cm} $r := r+1$ 
	\hfill{; update loop counter}
\item \hspace{1cm} Goto {\bf Main Loop} 
	\hfill{; try again}
\end{list}

%
%


%
%

This procedure for simulating $U$ has an intrinsic error probability
due to the possibility of failure in the gate teleportation procedure.
If we demand that the procedure succeed with probability one, then the
procedure may be repeated an arbitrarily large number of times, making
our simulation inefficient.  Fortunately, an efficient construction
can be devised, based on the threshold theorem for quantum
computation~\cite{Nielsen01a,Aharonov99a,Gottesman97a,Kitaev97a,Knill98a,Preskill98b}.
The idea is to use the measurement model to simulate a fault-tolerant
circuit in the standard model.  In such a fault-tolerant circuit any
element in the circuit can fail with some small probability $\epsilon
> 0$ (currently estimates put $\epsilon$ in the range $10^{-4}$ to
$10^{-6}$)), yet the circuit as a whole still succeeds with
probability arbitrarily close to one.  The single-qubit gates in the
fault-tolerant circuit can thus be simulated in the measurement model,
using at most $r$ iterations of the gate teleportation procedure,
where $(3/4)^r < \epsilon$.  Thus the total number of operations
required to simulate $U$ in the measurement model scales as
$O(\log(1/\epsilon))$.  If we assume a threshold of $10^{-5}$ then we
require $r = 41$ iterations to achieve a failure probability less than
$10^{-5}$.

%
%

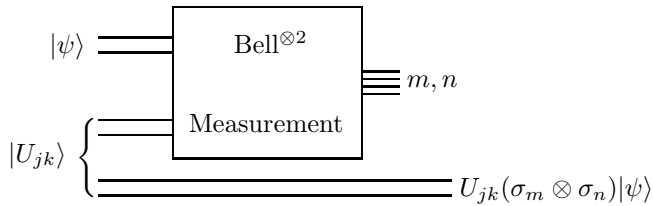
\begin{figure}[t]
\begin{center}
\setlength{\unitlength}{1cm}
\begin{picture}(7,4)
\put(0.15,2.9){$|\psi\rangle$}		
\put(-0.4,1.45){$|U_{jk}\rangle$}	
\put(0.5,1.42){$\Bigg\{$}		
\put(0.8,3.1){\line(1,0){1}}		
\put(0.8,2.9){\line(1,0){1}}		
\put(0.8,2){\line(1,0){1}}		
\put(0.8,1.8){\line(1,0){1}}		
\put(0.8,1.2){\line(1,0){4.7}}		
\put(0.8,1){\line(1,0){4.7}}		
\put(1.8,1.5){\framebox(2.5,2){}}	
\put(2.6,2.9){Bell$^{\otimes 2}$}	
\put(2,1.85){Measurement}		
\put(4.3,2.65){\line(1,0){0.5}}		
\put(4.3,2.55){\line(1,0){0.5}}		
\put(4.3,2.45){\line(1,0){0.5}}		
\put(4.3,2.35){\line(1,0){0.5}}		
\put(4.9,2.45){$m,n$}  			
\put(5.6,1){$U_{jk} (\sigma_m \otimes
	\sigma_n)|\psi\rangle$}		
\end{picture}
\end{center}

\caption{Circuit to teleport the two-qubit gate $U$ in the measurement
model of quantum computation.  The state $|U_{jk}\rangle$ is prepared
offline. Two Bell-basis measurements are done, one on the first and
third qubits, with outcome $m = 0,1,2,3$, and the second on the second
and fourth qubits, with outcome $n = 0,1,2,3$.  It will be useful
below to refer to this joint basis of the four qubits as the ``Bell
basis for four qubits''.  The output of the circuit is $U_{jk}
(\sigma_m \otimes \sigma_n) |\psi\rangle$, by a similar analysis to
the single-qubit case.\label{fig:two_qubits}} \end{figure}

Simulating two-qubit gates in the measurement model of quantum
computation is similar to simulation of single-qubit gates.  Let $U$
now denote an arbitrary {\em two}-qubit quantum gate.  To simulate $U$
we first define $|U_{jk}\rangle$ to be the result of applying $U
(\sigma_j \otimes \sigma_k)$ to the third and fourth qubits of
$(|00\rangle_{13}+|11\rangle_{13})\otimes
(|00\rangle_{24}+|11\rangle_{24}) /2$, where the subscripts label
which qubit is being referred to.  Such a state can be prepared by
first preparing the state $|0000\rangle$ and then measuring in the
orthonormal basis $|U_{jk}\rangle$.  Gate teleportation is now
achieved using Fig.~\ref{fig:two_qubits}, in a fashion analogous to
the single-qubit case.  Once again, the number of operations necessary
to achieve a failure probability at most $\epsilon > 0$ is
$O(\log(1/\epsilon))$ operations in the measurement model. 

%
%

We have shown that universal quantum computation can be performed
using measurements, quantum memory, and preparation of the $|0\rangle$
state.  This result appears rather mysterious --- after all, isn't it
the ability to evolve coherently that is most important to quantum
computation?  Some insight may be gained by noting that a similar
model is universal for classical computation.  Suppose we have the
following three abilities: (a) preparation of arbitrary probability
distributions over four bits; (b) storage of bits (memory); and (c)
measurements on up to four bits at a time.  To see that this model of
computation is universal for classical computation, suppose for
example that we want to simulate the {\sc not} gate on a single bit;
it will be clear how to generalize the result to arbitrary one- and
two-bit gates.  To do the simulation, first prepare two bits in the
state $(Y, \mbox{{\sc not}}(Y))$, where $Y$ is a classical random
variable which is zero with probability $1/2$ and one with probability
$1/2$.  To apply the {\sc not} gate to a classical bit $x$ we perform
a measurement of the first two bits of the string $(x,Y,\mbox{{\sc
not}}(Y))$.  This measurement has two outcomes, which correspond to
the case when $x$ is the same as $Y$, in which case the third bit
finishes in the state {\sc not}$(x)$, and the protocol succeeds, and
the case when $x$ is different to $Y$, in which case the protocol
fails, with the output of the third bit being $x$.  Fortunately, it is
possible to repeat the protocol so that with just a few repetitions we
obtain a high probability of successfully applying a {\sc not} gate.
It is straightforward to generalize this construction to any classical
gate, and thus to perform universal classical computation using the
preparation of arbitrary probability distributions over four bits,
memory, and measurements on up to four bits at a time.

%
%

What effect does experimental noise have on the measurement model of
quantum computation?  It is straightforward to prove that an analogue
of the threshold theorem for the standard model of quantum
computation~\cite{Nielsen01a,Aharonov99a,Gottesman97a,Kitaev97a,Knill98a,Preskill98b}
holds in the measurement model.  In brief, the idea is that for any
circuit in the measurement model, we can construct an equivalent
circuit in the standard model of quantum computation, then make that
circuit fault-tolerant using well-known procedures, and finally obtain
a fault-tolerant circuit in the measurement model by simulating the
fault-tolerant circuit in the standard model.  This procedure for
obtaining fault-tolerant circuits in the measurement model is rather
indirect, and it would be interesting to obtain a fault-tolerant
construction more directly, and to compare the threshold thus obtained
to thresholds in the standard model of quantum computation.

Does the measurement model of quantum computation have any practical
implications as a means of simplifying experimental proposals for
quantum computation?  Unfortunately, the four-qubit projective
measurements used in the present construction are not easily
implemented in most physical systems, and thus the construction is
mainly of theoretical interest.  However, it may be possible to
simplify the measurements sufficiently that the model is of practical
interest, ideally by reducing the number of systems involved in the
measurements being performed.

%
%

We have shown that universal quantum computation can be performed
using projective measurement, quantum memory, and the ability to
prepare the state $|0\rangle$.  This result emphasizes the power that
comes from the ability to perform measurements during a computation.
The simplicity of the model may also make it useful for the study of
quantum computational complexity.  Finally, by showing that coherent
dynamical operations are not necessary for universal quantum
computation, the model gives valuable insight into the fundamental
question of what gives quantum computers their power.

\section*{Acknowledgments} This paper was stimulated by a discussion
with Damian~Pope about what makes quantum computers powerful.  Thanks
to Michael~Bremner, Ike~Chuang, Jennifer~Dodd, Alexei~Gilchrist,
Tobias~Osborne, and Damian~Pope for a thorough reading of the
manuscript.


\end{multicols}

\end{document}